\documentclass[aps,twocolumn,prl,floatfix,superscriptaddress,showpacs,longbilbiography]{revtex4-1}
\usepackage[utf8]{inputenc}

\usepackage{graphicx}
\usepackage{dcolumn}
\usepackage{bm}

\usepackage{braket}
\usepackage{amsfonts}
\usepackage{amsmath}
\usepackage{amssymb}

\usepackage{color,soul}
\usepackage{wasysym}
\usepackage{mathrsfs}
\usepackage{times}

\usepackage[dvipsnames,svgnames,table]{xcolor}
\usepackage{hyperref}
\usepackage{fancyhdr}
\usepackage{float}
\usepackage[english]{babel}
\usepackage[caption=false]{subfig}
\usepackage{braket}

\hypersetup{
pdfstartview={FitH},
colorlinks=true,    
linkcolor=NavyBlue, 
citecolor=Maroon,   
filecolor=NavyBlue, 
urlcolor=NavyBlue   
}

\newcommand{\bea}{\begin{eqnarray}}
\newcommand{\eea}{\end{eqnarray}}

\newcommand{\overbar}[1]{\mkern 1.5mu\overline{\mkern-1.5mu#1\mkern-1.5mu}\mkern 1.5mu}

\begin{document}
\title{Topological Signatures in Quantum Transport in Anomalous Floquet-Anderson Insulators}

\author{Esteban A. Rodríguez-Mena}
\author{L. E. F. Foa Torres}
\affiliation{Departamento de F\'{\i}sica, Facultad de Ciencias F\'{\i}sicas y Matem\'aticas, Universidad de Chile, Santiago, Chile}

\begin{abstract}
Topological states require the presence of extended bulk states, as usually found in the picture of energy bands and topological states bridging the bulk gaps. But in driven systems this can be circumvented, and one can get topological states coexisting with fully localized bulk states, as in the case of the anomalous Floquet-Anderson insulator. Here, we show the fingerprints of this peculiar topological phase in the transport properties and their dependence on the disorder strength, geometrical configuration (two-terminal and multiterminal setups) and details of the driving protocol.

\end{abstract}

\date{\today}
\maketitle

\section{Introduction} 

Topological arguments have reshaped condensed matter physics~\cite{noauthor_topology_2016,castelvecchi_strange_2017,asboth_short_2016}. The search for topological states has expanded from materials in different dimensions~\cite{hasan_colloquium_2010,xiao_berry_2010,konig_quantum_2007,hsieh_topological_2008,bradlyn_topological_2017,zhang_catalogue_2019} to fertile new grounds including ultracold matter~\cite{atala_direct_2013,jotzu_experimental_2014,eckardt_colloquium:_2017}, photonic~\cite{ozawa_topological_2019} and acoustic systems~\cite{yang_topological_2015,esmann_topological_2018}, and also to new mechanisms, from dissipation~\cite{budich_dissipative_2015} and non-Hermitian terms~\cite{martinez_alvarez_topological_2018,foa_torres_perspective_2019,gong_topological_2018,kunst_biorthogonal_2018}, to time-dependent potentials~\footnote{Here we should count both external fields and self-induced ones as in Ref.~\cite{rudner_self-induced_2019} which presents a \textit{self-Floquet} phenomenon.} as in Floquet topological insulators~\cite{oka_photovoltaic_2009,lindner_floquet_2011,rudner_self-induced_2019}. As usual in a quest for new terrain, our curiosity aims at unveiling what is truly distinctive and predicting hallmarks that allow to evidence it.

The anomalous Floquet-Anderson insulator (AFAI)~\cite{titum_anomalous_2016} is a topological phase unique to driven systems in two-dimensions and which critically relies on disorder, another research axis with strong current interest~\cite{varjas_computation_2019}. The AFAI is one of the latest members of the larger family of the so-called Floquet topological insulators~\cite{oka_photovoltaic_2009,lindner_floquet_2011,rudner_floquet_2019} spanning systems in all dimensions~\cite{gomez-leon_floquet-bloch_2013,asboth_chiral_2014,fedorova_cherpakova_limits_2019,calvo_floquet_2015,dauphin_loading_2017}. Its defining feature being the presence of robust topological states (in a two-dimensional system) which persist even when all bulk states are localized by disorder. This has no counterpart in static systems where the existence of such chiral edge states requires the presence of delocalized states so as to stop the spectral flow~\cite{halperin_quantized_1982}.

By elegantly exploiting a characteristic of Floquet systems, Titum and collaborators~\cite{titum_anomalous_2016} have found an exception to Halperin's argument~\cite{halperin_quantized_1982}: since the quasienergies can be chosen on a torus, the edge states can be present throughout the full span of the quasienergy Brillouin zone warping around it. The nature of such states and the transitions from the pristine to the (bulk) Anderson localized phase may guard fascinating insights and has motivated theoretical proposals~\cite{nathan_quantized_2017} and experiments~\cite{mukherjee_experimental_2017,maczewsky_observation_2017}. But what are the fingerprints of the AFAI in transport experiments? In spite of the interest in transport of Floquet topological states both theoretically~\cite{kundu_transport_2013,kundu_effective_2014,foa_torres_multiterminal_2014,dehghani_out--equilibrium_2015,farrell_photon-inhibited_2015,dehghani_occupation_2016,fruchart_probing_2016,atteia_ballistic_2017,peralta_gavensky_time-resolved_2018,balabanov_transport_2019,sato_microscopic_2019,sato_light-induced_2019} and experimentally~\cite{mciver_light-induced_2018}, much less is known for the case of the AFAI. The only work~\cite{kundu_quantized_2017} on this phase focused on the high bias voltage regime in a two-terminal configuration, finding a quantized charge pumping~\cite{kundu_quantized_2017}.

Here we address the transport response of the AFAI in two-terminal and multiterminal setups from where we can obtain further insight on this new phase. Notably, we can visualize the transition from the ordered phase to the Anderson localized phase where robust transport is observed for energies throughout the full quasienergy spectrum, thereby erasing the usual frontier between energies within or outside the gap. This is shown to be detectable both in two and three terminal measurements. We also examine the pumping contribution which appears because of the interplay of driving and the (unavoidable) symmetry breaking due to disorder. Our results prepare the way for new experiments aimed at this exotic topological phase.

\begin{figure}
\centering
\includegraphics[width=1.0\linewidth]{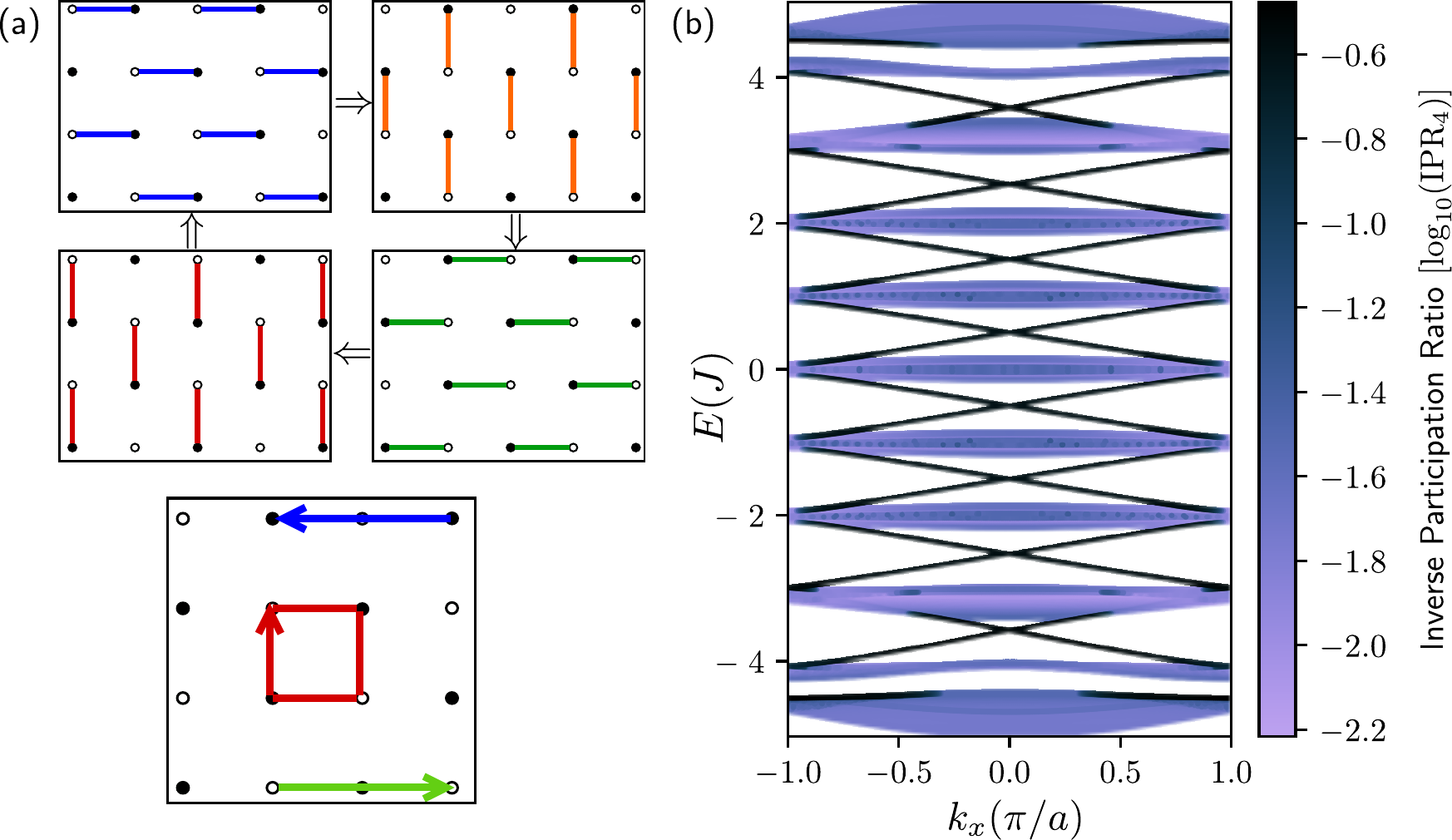}
\caption{(a) Driving protocol in a model of the anomalous Floquet-Anderson insulator in a square lattice. The hoppings (full lines connecting the sites represented by circles) are turned on and off in a sequence as indicated in the four different panels. The magnitude of the hopping matrix elements, $J$, is chosen so that the probability to hop during the on-interval is equal to unity. The driving protocol produces chiral edge states at the boundaries of the ribbon and no effect in the bulk. (b) Quasienergy band structure in terms of crystal momentum in the first Brillouin zone. The calculations considered replicas from -4 to +4. The color scale encodes the inverse participation ratio in log scale showing the strong localization of the chiral modes that bridge the gap. The states in the bands remain delocalized.}
\label{fig1}
\end{figure}

\section{Hamiltonian model for the anomalous Floquet-Anderson Insulator (AFAI)}
We use the Floquet-Anderson insulator model introduced in~\cite{titum_anomalous_2016}. A square lattice tight-binding model with A and B-type sites (represented in Fig.~\ref{fig1}(a) with black and white circles), is driven so that the hoppings between sites are turned-on and off according to a four step cycle of period $T$ depicted in Fig.~\ref{fig1}(a) (a particle starting at any site would describe a loop counterclockwise). This is represented through a time-dependent contribution $\mathcal{H}_{\text{hop}}(t)$ to the full Hamiltonian. The other two ingredients are: a static staggering potential with strength $\Delta$, and a static on-site Anderson disorder. The Hamiltonian reads:

\begin{equation}
\mathcal{H}_{\text{AFAI}}(t)=\mathcal{H}_{\text{hop}}(t)+\sum_{\mathbf{r}} (E_\mathbf{r}+V_{\mathbf{r}}) c^{\dagger}_{\mathbf{r}}c^{}_{\mathbf{r}}
\end{equation}
where $c^{\dagger}_{\mathbf{r}}$ and $c^{}_{\mathbf{r}}$ are the fermionic creation and annihilation operators at site $\mathbf{r}$, $E_\mathbf{r}$ takes the value $\Delta$ ($-\Delta$) for A-sites (B-sites) thus representing the staggering potential, and $V_{\mathbf{r}}$ is a random potential uniformly distributed in $[-\delta V, \delta V]$ with $dS\equiv \delta V T$ the \emph{disorder strength}. The time-dependent term $\mathcal{H}_{\text{hop}}(t)$ contains the piece-wise constant nearest-neighbors hopping terms with magnitude $J$. $J$ is chosen so that a particle starting at one site can get fully transferred to the next site at the proper interval, $JT/4=\pi/2$. All our simulations in this paper use the Kwant package~\cite{groth_kwant:_2014}.

The Fourier components of each of the time-dependent hoppings $j^{(a)}(t)$ in $\mathcal{H}_{\text{hop}}(t)$ are:
\begin{align}
j^{(a)}(t) & ={\displaystyle \sum_{m}} \, j^{(a)}_{m}e^{-im\Omega t},\\
j_{m}^{(a)} & =\begin{cases}
\frac{J}{i2\pi m}e^{im2\pi a}\left(e^{\frac{i2\pi m}{4}}-1\right) & \forall m\neq0\\
J/4 & m=0
\end{cases}.
\end{align}
The superscript $(a)$ denotes the stage of the driving protocol, $a \in 1,...,4$.

\textit{Floquet theory for the spectral and transport properties.--} We use Floquet theory~\cite{kohler_driven_2005,platero_photon-assisted_2004,shirley_solution_1965} to solve for the spectral and transport properties. In the following we outline the main steps of our calculations.

For a time-periodic Hamiltonian $\mathcal{H}(t)=\mathcal{H}(t+T)$, the Floquet theorem assures that there is a complete set of solutions of the form $\psi_{\alpha}(\mathbf{r},t)=\exp(-i\varepsilon_{\alpha} t/\hbar) \phi_{\alpha}(\mathbf{r},t)$, where $\varepsilon_{\alpha}$ are the quasienergies and $\phi_{\alpha}(\mathbf{r},t+T)=\phi_{\alpha}(\mathbf{r},t)$ are the Floquet states obeying: $\mathcal{H}_F \phi_{\alpha} (\mathbf{r},t)=\varepsilon_{\alpha} \phi_{\alpha}(\mathbf{r},t)$ where $\mathcal{H}_F \equiv \mathcal{H}-i\hbar \partial/\partial t$ is the Floquet Hamiltonian. Thus, one gets an eigenvalue problem in the direct product space (Floquet or Sambe space~\cite{sambe_steady_1973}) $\mathcal{R} \otimes \mathcal{T}$ where $\mathcal{R}$ is the usual Hilbert space and $\mathcal{T}$ the space of $T-$periodic functions spanned by $\braket{t | n}=\exp{(i n \Omega t)}$ (the index $n$ is often called the replica index). The matrix elements of $\mathcal{H}_F$ are given by:

\begin{multline}
\bra{\mathbf{r},n}\mathcal{H}_{F}\ket{\mathbf{r}',m} = \frac{1}{T} \int^{T}_{0} \bra{\mathbf{r}}\mathcal{H}(\mathbf{r},t)\ket{\mathbf{r}'}e^{-i(n-m)\Omega t} dt + \\
+m\hbar \Omega \delta_{n,m} \delta_{\mathbf{r},\mathbf{r}'}.
\end{multline}
The number of Floquet replicas is truncated to ensure convergence of the desired physical quantity. This leads to a solution which has two main features which are desirable for our problem: it is variational (and hence non-perturbative~\cite{moskalets_scattering_2011,calvo_non-perturbative_2013}), in the sense that the result can be improved by keeping a larger portion of the full state space when truncating; and it is valid in the non-adiabatic frequency regime. Thus, in Floquet space the driving term provides hopping amplitudes to otherwise disconnected sites with a staggered onsite energy. These hopping matrix elements couple different replicas with an amplitude given by the Fourier components noted earlier.

The transport calculations are based on Floquet scattering theory~\cite{moskalets_floquet_2002,camalet_current_2003,kohler_driven_2005, moskalets_scattering_2011}, which has been used for different materials and devices~\cite{kundu_effective_2014,foa_torres_controlling_2009}. In a non-interacting case such as this one it is equivalent to the Keldysh formalism~\cite{arrachea_relation_2006}. The sample described by the Hamiltonian $\mathcal{H}_{\text{AFAI}}(t)$ is now connected to leads or terminals (in what follows we will consider two and three terminal configurations). Since we are not interested in the details of the leads we will use leads with a square lattice structure and nearest-neighbors hoppings, the onsite energies and hoppings are tuned as to minimize reflections at the boundary with the driven sample.

The time-averaged current at lead $\alpha$ is given by~\cite{moskalets_floquet_2002,kohler_driven_2005,moskalets_scattering_2011}:

\begin{equation}
\label{Floquet-Current}
\mathcal{\overbar{I}}_{\alpha} = \frac{2e^2}{h} \sum_{\beta \neq \alpha} \sum_n \displaystyle{\int} \left [\mathcal{T}^{(n)}_{\beta,\alpha}(\varepsilon)f_{\alpha}(\varepsilon) - \mathcal{T}^{(n)}_{\alpha,\beta}(\varepsilon)f_{\beta}(\varepsilon) \right ] d\varepsilon,
\end{equation}

\noindent where $\mathcal{T}^{(n)}_{\beta, \alpha} (E)$ is the transmission probability for an electron coming from lead $\alpha$ with energy $E$ to exit at lead $\beta$ having emitted $n$ photons, and $f_{\alpha}$ are the Fermi-Dirac distributions at lead $\alpha$ which are kept in equilibrium. The previous equation can also be written in terms of the total transmission probabilities:

\begin{equation}
\mathcal{T}_{\beta, \alpha} = \sum_{n} \mathcal{T}^{(n)}_{\beta, \alpha} (\varepsilon).
\end{equation}

In the case of two-terminals $\alpha,\beta=L,R$ (left, right), unitarity enforces the relation $\overbar{\cal{I}}_L=-\overbar{\cal{I}}_R\equiv\overbar{\cal{I}}$. Eq. (\ref{Floquet-Current}) can be written as:

\begin{equation}
\overbar{{\cal I}}=\frac{2e}{h}\!\int\!\Big[{\cal T}(\varepsilon) [f_{L}(\varepsilon)-f_{R}(\varepsilon)]+\delta {\cal T}^{}(\varepsilon) [f_{L}(\varepsilon)+f_{R}(\varepsilon)]\Big]  d\varepsilon,
\label{Floquet-Current-2-terminals}
\end{equation}
where $\mathcal{T}(\varepsilon)=[{\cal T}_{R,L}(\varepsilon) + {\cal T}_{L,R}(\varepsilon)]/2$ and $\delta {\cal T} = ({\cal T}_{R,L} - {\cal T}_{L,R})/2$ characterizes the average transmission probability and the asymmetry between left and right. This asymmetry gives rise to a current even at zero bias voltage. Such \textit{pumping} currents have intriguing consequences in contexts ranging from transport in conductors  modulated by gate voltages~\cite{kaestner_non-adiabatic_2015,giazotto_josephson_2011,moskalets_floquet_2002} to shift photocurrents~\cite{bajpai_spatio-temporal_2019}. In the zero temperature limit and to linear order in the applied bias voltage $V$ one obtains~\cite{foa_torres_multiterminal_2014}:

\begin{equation}
\overbar{{\cal I}}\simeq \frac{2e^2}{h} {\cal T}(\varepsilon_F) \times V + \frac{4e}{h} \int \delta {\cal T}(\varepsilon)f(\varepsilon)  d\varepsilon.
\label{Floquet-Current-2-terminals-zerobias}
\end{equation}

\noindent The second term represents a \textit{pumping current} originated by the asymmetry of the transmission coefficients and does not depend on the (small) bias voltage. Hence, the differential conductance $G\equiv\frac{d\overbar{{\cal I}}}{dV}$ is given by the average of the transmission probabilities from left to right \textit{and} from right to left~\cite{foa_torres_multiterminal_2014}: $(2e^2/h){\cal T}(\varepsilon_F)$.

\begin{figure}
\centering
\includegraphics[width=0.9\linewidth]{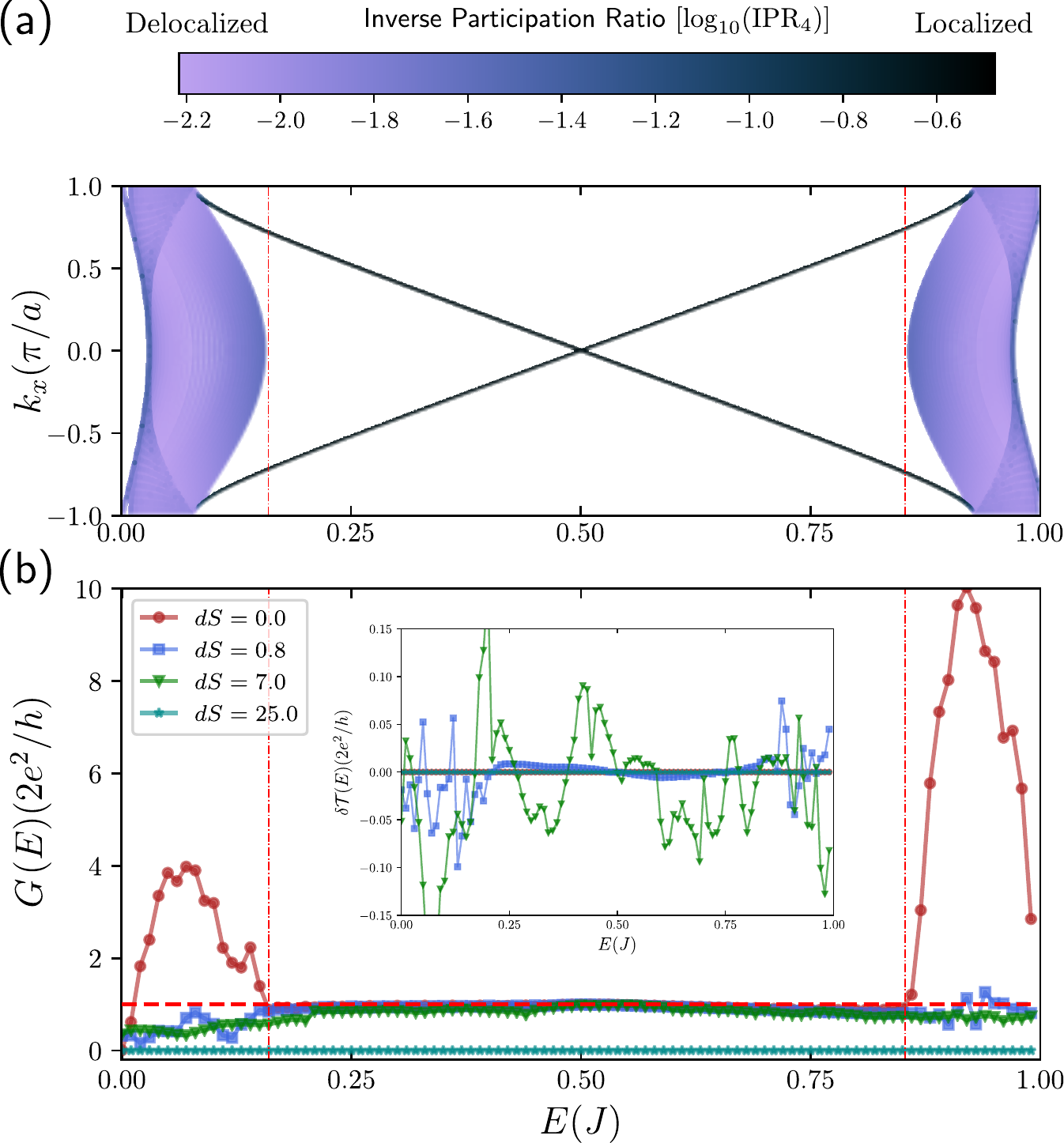}
\caption{(a) Quasienergy as a function of the wave-vector along the translationally invariant direction in a ribbon geometry for the AFAI model used in the text. Our study of transport properties is focused in the first non-trivial (region denoted between dashed vertical red lines) gap which is bridged by two counterpropagating states. (b) Two terminal differential conductance as a function the Fermi energy in the undriven leads. The conductance reaches values close to the conductance quantum $2e^2/h$ for energies within the non-trivial gap.  The two terminal conductance in the bands is suppressed as the disorder intensity increases until the anomalous Floquet Anderson insulator regime sets in and the conductance saturates slightly below the theoretical limit of  $2e^2/h$. For high disorder intensities we reach the strong localization regime. The inset of panel (b) shows the transmission asymmetry $\delta {\cal T} = ({\cal T}_{R,L} - {\cal T}_{L,R})/2$ which is responsible of pumped current. The calculations correspond to $W=40$ unit cells and the lead parameters are $J_{\text{lead}}=0.258$ and $E_{\text{lead}}=0.528$.}
\label{fig2}
\end{figure}

\section{Spectral properties of the pristine model.} 
To motivate our forthcoming discussion we start by looking at the spectral properties of the model defined earlier without the disorder term.  The full Floquet spectrum for a pristine ribbon is shown in Fig.~\ref{fig1}(b). The color scale encodes information on the localization of each state as given by the \emph{inverse participation ratio} (IPR)~\cite{kramer_localization:_1993,evers_anderson_2008}:
\begin{equation}
\text{IPR} \equiv \sum_{\mathbf{r}} |\psi(\mathbf{r})|^{4}.
\end{equation}
For the edge states, the inverse of the IPR is a measure of the average spread $\xi$ (perpendicular to ribbon edge), while for extended bulk states the inverse IPR is proportional to $\xi^d$ ($d$ is the spatial dimension). Thus, the color scale goes from fully extended (light color) to localized (dark color). The topological edge states are chiral and bridge the bulk gap containing extended states (Fig.~\ref{fig2}(a) shows a zoom of the Floquet bandstructure around one of the gaps). Up to this point, the topological states encounter extended bulk states once they reach the bands, as usual. As we will see later on, the states in the bands localize when the disorder reaches a threshold. In contrast, the topological states withstand moderate to high disorder and, above the threshold, they span the entire Floquet-Brillouin zone. 

\section{Conductance in a two-terminal setup} 
Let us now focus on the transport properties. A sample of length $L$ and width $W$ is connected to simple (undriven) electrodes as explained earlier. These electrodes consist of a square lattice with the same geometry as the sample, with onsite energies ($E_{\text{lead}}$) and nearest neighbors hopping ($J_{\text{lead}}$). $E_{\text{lead}}$ and $J_{\text{lead}}$ are chosen as to minimize the reflections at the driven-undriven interphase (as we want to probe the properties of the driven region).

Figure~\ref{fig2}(b) shows the two-terminal conductance (defined from Eq. (\ref{Floquet-Current-2-terminals})) as a function of the Fermi energy in the leads. Each curve corresponds to a value of the disorder strength. In the energy region corresponding to the bulk gap (marked with dashed red lines in Fig.~\ref{fig2}(a)) the conductance has a plateau at $(0.94\pm0.04)2e^2/h$, $(0.92\pm0.05)2e^2/h$, $(0.84\pm0.08)2e^2/h$ for $dS=0.0$, $0.8$, and $7$, respectively. These values are close to the theoretical limit of $2e^2/h$ expected for one ballistic channel. The differences are attributed to the still imperfect matching at the driven-undriven interface, a problem which has been reported already for other driven systems in a scattering setup~\cite{kundu_effective_2014,foa_torres_multiterminal_2014}. As the disorder increases, the conductance plateau remains almost unaltered (one of the fingerprints of topological states) until very high values of disorder when the conductance falls to zero. 

\begin{figure*}[hbt!]
\centering
\includegraphics[width=0.9\linewidth]{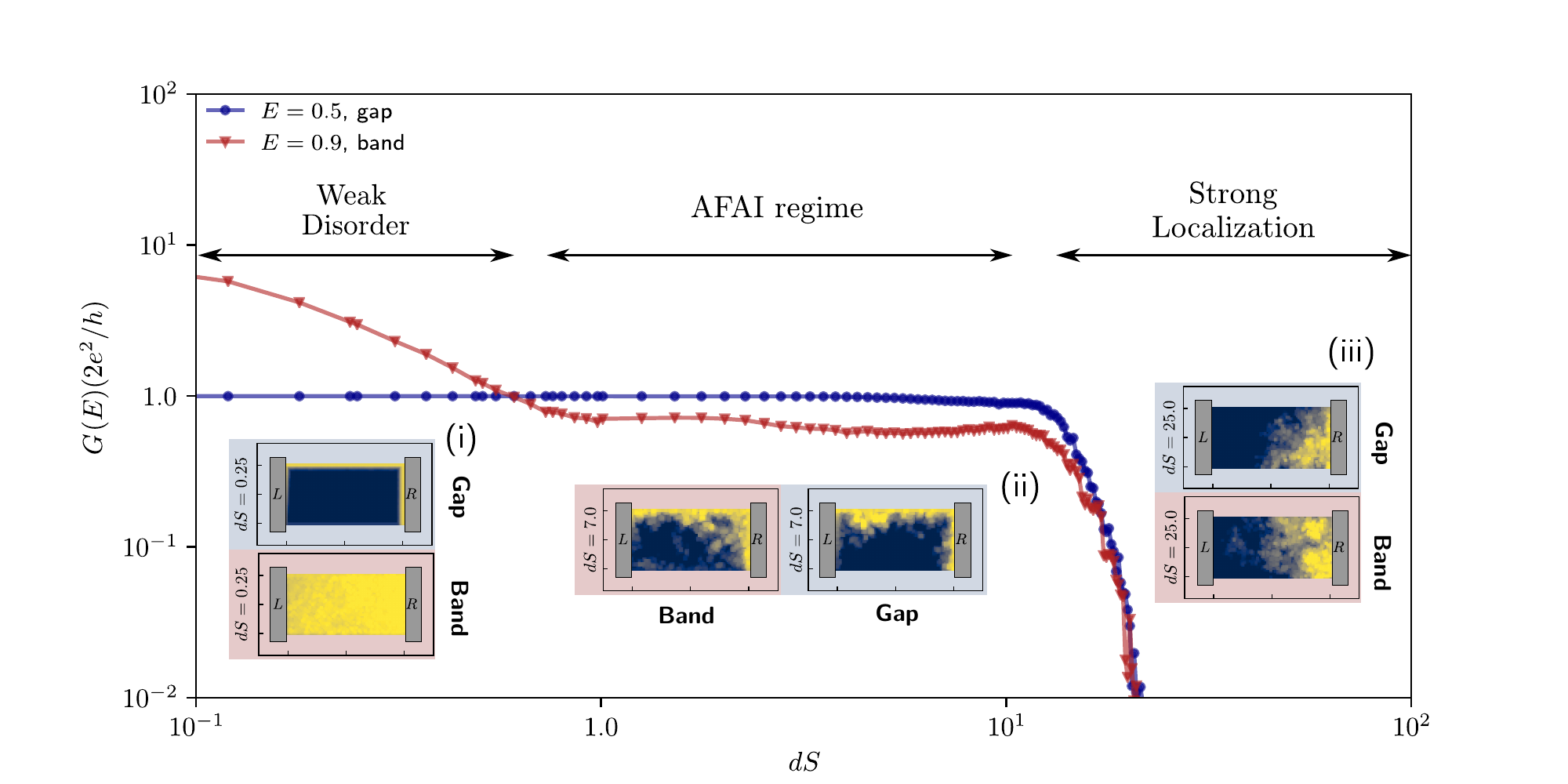}
\caption{Two terminal conductance variation in terms of the disorder strength $dS$. In red we show for an energy value within the band and blue for an energy inside the gap. Conductance probes to be constant for moderate disorder strength (region labeled as \emph{AFAI regime}) both in gap as in band energy regions. For intense disorder trivial strong localization manifests as a high suppression of the conductance. Computation was performed over an average of fifty disorder realizations.}
\label{fig3}
\end{figure*}

But unlike other topological phases, the most interesting features of the AFAI appear within the energy range of the bulk bands, in the outer regions of the energy scale shown in Fig.~\ref{fig2}(b). The conductance goes down rapidly as the disorder increases reaching a plateau close to the quantum limit of $2e^2/h$ for $dS\gtrsim1$. This plateau remains up to $dS\gtrsim12$ when the topological phase finally breaks down. Figure~\ref{fig3} shows the two-terminal conductance as a function of disorder for $E=0.5$ (i.e. within the bulk gap) and $E=0.9$ (i.e. within a band). The transitions are marked in the plot. These two transition points witness first, the phase transition from the pristine system to the AFAI phase, and second, from the AFAI to the strongly localized regime.

The insets in Fig.~\ref{fig3} show the scattering wavefunctions at selected values of disorder (the incident condition is from the right). In spite of the contrasting behavior at weak disorder (Fig.~\ref{fig3}(i)), the scattering wavefunctions become quite similar for moderate disorder, thereby showing a fingerprint of the AFAI regime. The smaller conductance values for $E=0.9$ (energy within the bulk bands) could be originated either from weak channels bridging the edges created by the localized states or a higher reflection at the contacts. Close examination of our full numerical results in the three terminal setup indicate that the second scenario is in place.
When the disorder is strong enough, localization sets in to impede transport as evidenced in Fig.~\ref{fig3}(ii), irrespective of the energy.

Another feature that is already visible in this two-terminal case is the existence of pumping currents~\cite{moskalets_floquet_2002,kohler_driven_2005}, produced by the asymmetry of the transmission coefficients ${\cal T}_{R,L}(\varepsilon) \neq {\cal T}_{R,L}(\varepsilon)$ which lead to a finite current even at zero bias voltage. Although our definition of the conductance $G$ takes into account this asymmetry, it is interesting to see how large it is. The asymmetry $\delta {\cal T} = ({\cal T}_{R,L} - {\cal T}_{L,R})/2$ is shown in the inset of Fig.~\ref{fig2}(b) for the same values of disorder as in the main panel. While for the pristine system there is no pumping (as the two leads are equivalent), this changes as soon as disorder is introduced and cannot be neglected.

\begin{figure}[ht!]
\includegraphics[width=0.9\linewidth]{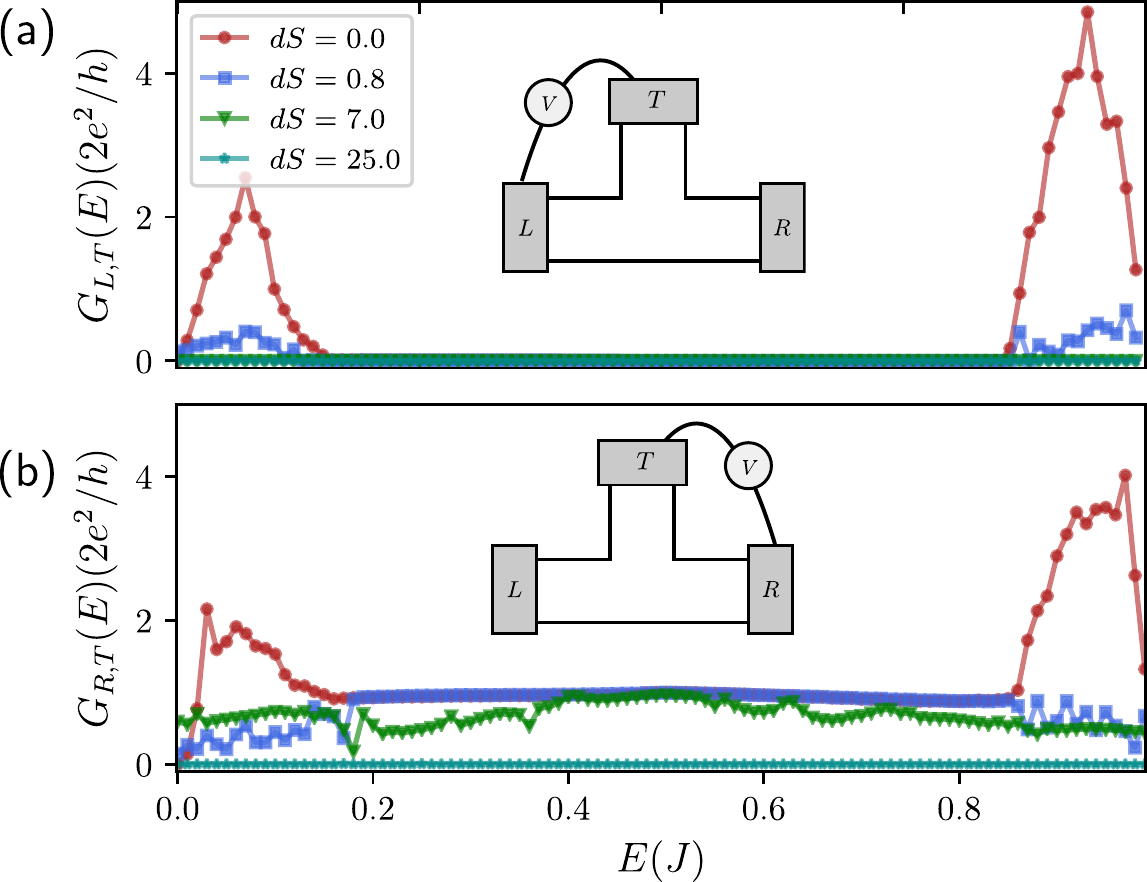}
\caption{Three terminal differential conductances for a T-shaped system as function of energy. Depending on the energy of the incident modes, the chirality expected from the anomalous Floquet Anderson states develops. The onset of the anomalous Floquet-Anderson regime is most clearly seen for energies within the bulk bands and $dS=7.0$.}
\label{fig4}
\end{figure}

\begin{figure*}[ht!]
\includegraphics[width=0.7\linewidth]{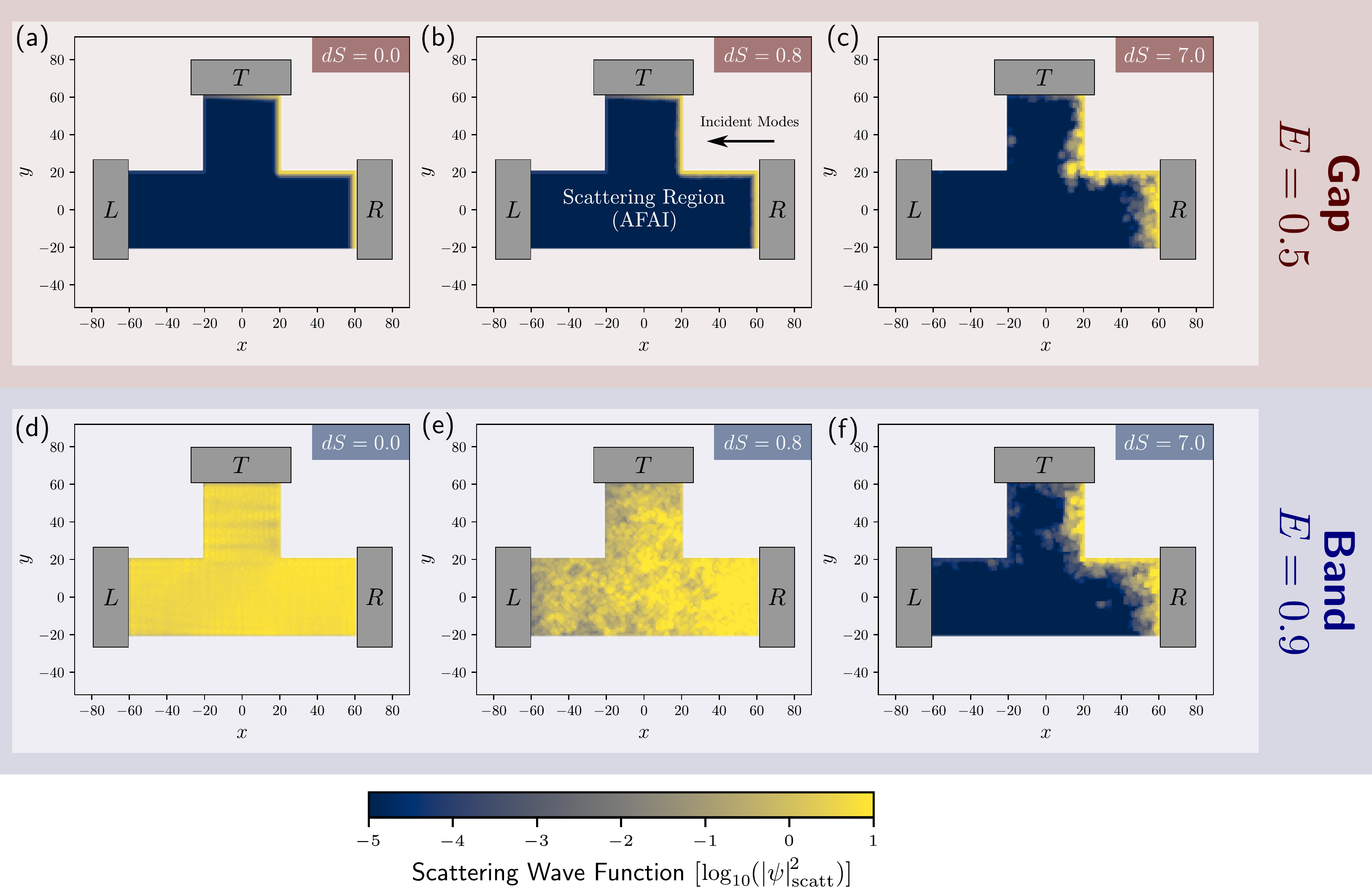}
\caption{Scattering wave functions for three terminal AFAI device for two energies, $E=0.5$ (within the bulk gap) and $E=0.9$ (within the bulk bands) for different values of disorder. As disorder is increased,  the bulk states localize until a single chiral state remains (panel f), thereby indicating the anomalous Floquet-Anderson regime.}
\label{fig5}
\end{figure*}

\section{Conductance in multiterminal setup} 

Let us now turn to the search of further topological signatures in transport, this time in a multiterminal setup. The Hall response is the quintessential feature of chiral topological states, as typically found in the integer quantum Hall effect~\cite{von_klitzing_new_1980}. In our case, the numerical exploration of the Hall resistance is cumbersome because of the existence of a non-zero response at zero bias which would require to make a self-consistent procedure to ensure that the current at the voltmeter vanishes \footnote{Note that in the presence of disorder one cannot impose a symmetry condition assuring that no pumping occurs as in~\cite{foa_torres_multiterminal_2014}.}. Instead, here follow a different alternative~\cite{fruchart_probing_2016}, circumventing the complications introduced by the pumping contributions. 

Let us define the differential conductance as the sensitivity of the current entering the lead $\alpha$ to variations of the chemical potential $\mu_{\beta}$ of the lead $\beta$ as proposed in~\cite{fruchart_probing_2016}:

\begin{equation}
G_{\alpha, \beta} (\varepsilon_F)\equiv -e \frac{d  \overbar{\cal I}_{\alpha}}{d\mu_{\beta}}= \frac{2e^2}{h} \mathcal{T}_{\alpha, \beta} (\varepsilon_{F})
\end{equation}
where the last equality follows from the low bias and low temperature limit. We stress that this definition of differential conductance is 
not symmetric in the various chemical potentials. 

The chirality of the states in the AFAI model can be appreciated by comparing $G_{L,T}$ and $G_{R,T}$ ($L$, $T$ and $R$ stand for left, top and right contacts as shown in the insets of Fig.~\ref{fig4}), the differential conductances, in a setup with three terminals as shown in Fig.~\ref{fig4}(a) and (b) as a function of the Fermi energy. For vanishing to moderate disorder ($dS=0$ and $dS=0.8$) and within the bulk gap region, the transmission is almost perfect from $T$ to $R$ while it vanishes from $T$ to $L$. For larger $dS=7.0$ $G_{R,T}$ degrades slightly within the gap region but improves within the first band ($G_{L,T}$ almost vanishes in this range), indicating that bulk states have become Anderson localized. This is the crucial trait of the anomalous Floquet Anderson regime.

Deviations from the perfect behavior, this is, deviations from one conductance quantum at $dS=7.0$ are attributed to imperfect matching at the sample-lead interface. This offers a more realistic picture of what could be seen in experiments in optical or photonic systems.

Figure~\ref{fig5} provides further insight on the chirality of the edge states and the transition from the weak localization to the AFAI regime. Panels (a-c) and (d-f) show the scattering states for $E=0.5$ and $E=0.9$) with incidence from the right side. The onset of the AFAI regime becomes clear for $dS=7.0$ in agreement with our discussion of Fig.~\ref{fig4}. Although the scattering wavefunctions are beyond our reach in condensed matter experiments, they could be probed in photonic systems or ultracold matter. 

\section{Final remarks} 

We have presented a study of transport in the anomalous Floquet-Anderson insulator, a topological phase unique to driven systems. Our results show the signatures as a function of disorder for a finite sample realizing this model. The onset of a regime where robust edge states coexist with an Anderson localized bulk is evidenced in both the two-terminal and three-terminal conductance in the linear response regime. Another key feature, the chirality of the edge states, manifests more clearly in a three-terminal setup. Further evidence can be obtained from the visualization of the scattering states in real-space, something that could be accessible in photonic experiments. We hope that our results foster the experimental search of these new phases.

\begin{acknowledgments}
EARM acknowledges support from CONICYT (Chile) national doctoral grant and Matías Berdakin, Lucila Peralta and Gonzalo Usaj the fruitful discussions. We thank the support of FondeCyT (Chile) under grant number 1170917.
\end{acknowledgments}

\end{document}